# *Symmetries in evolving space time from present to prior universes*

Andrew Beckwith

abeckwith@UH.edu

## Abstract

We wish to demonstrate how the presence of a dynamic cosmological parameter and a model of the universe as a quantum computer can be combined to give us a picture of early universe inflationary physics which is quite non linear, with a distinct bifurcation. We claim that the bifurcation is linked to conditions which falsify initial conditions for casually continuous space times in the bridge between pre inflation to inflationary cosmology. And also give necessary and sufficient conditions for the existence of a wormhole bridge between a prior contracting universe, and our present universe via a time dependent version of the Wheeler DeWitt equation. The existence of the solution of the Wheeler DeWitt equation with a pseudo time like component provides an additional symmetry to space time evolution which is broken by the chaotic regime of the scale factor, leading to a bifurcation in the evolution of the quintessence scalar field. The causal discontinuity is a direct outgrowth of a discrete time treatment of the Friedmann equation and lack of causal set ordering on a discrete early universe geometry as presented in Fay Dowker's set theoretic analysis of quantum gravity phenomenology

# Introduction.

We explicitly use a worm hole solution of the Wheeler DeWitt equation as proving a thermal bridge between a prior universe which is contracting to a singularity, to the initial phases of our present universe, to provide a time dependent symmetry which is broken via casual discontinuity. The casual discontinuity is presented by using Fay Dowker's characterization of causal sets [1], and a discrete interval version of the Friedmann equation. In addition, we use the datum Fay Dowker brings up about initial phases of the universe as comprising discrete set entries as arguing this forms a conceptual bridge between loop quantum gravity [2], the Wheeler- DeWitt equation [3], and Brane world physics [4]

To give us a start, a) Loop quantum gravity may be giving us a template of thermal input which affirms that relic graviton production can exist. B) Brane world models as constructed by Randall [5] and Sundrum permit the low entropy conditions Carroll and Chen predicted in 2005 [6] as well as how transition from the brane world model to be consistent with the 10 to the 32 Kelvin conditions stated by Weinberg in 1972 [7] as necessary for quantum gravity . From that there is a transition to Guth style inflation.

We will organize how to get to these bench marks via the following steps After we present the Wormhole solution of the Wheeler DeWitt equation, we 2$^{nd}$ form an argument for causal discontinuity via using Fay Dowker's characterization of discrete elements of a causal set and tie that in with causal discontinuity we claim is given by a discrete version of the Friedmann equation for expansion of the scalar factor.. 3$^{rd}$, we will outline Seth Lloyd's model of the universe as a quantum computer.[8] 4$^{th}$ , we will bring up the vacuum energy as a temperature dependent cosmological parameter, varying from the moment of nucleation to a phase transition which involves axion (or in SUSY parlance, saxion) wall formation and decay, as a contribution to dark energy [9].5$^{th}$ we bring up relic graviton burst behavior in the onset of a transfer of thermally based vacuum energy[10,11] 6$^{th}$, we discuss how the bifurcation of the scale factor occurs, in tandem with the short term behavior of a quintessence field we claim is short lived, and contributes, when decaying to real time values as a result of a large vacuum energy to the creating of dark energy later on.

The early universe behavior of the scalar quintessence field is next outlined. We claim that it disappears at an early date, i.e. well before the onset of the CMBR (i.e. 400 thousand years after the big bang), in such a way as to preserve the symmetry of the quintessence field so outlined by traditional inflation, but at an earlier date. This leads to a bifurcation of the scale factor $a(t)$ between an initial exponential expansion , to a chaotic regime, reflecting upon casual discontinuity, then finally to the present to future cosmological conditions leading to acceleration of the expansion of the universe under today's condition.[12]

## I. Presenting the first symmetry, i.e. the Wheeler De Witt worm hole transferal of vacuum energy from a prior universe to today's universe

We claim that a prior universe is linked via a worm hole solution of the Wheeler De Witt equation to a present universe configuration, due to a pseudo time component to the Wheeler De Witt equation. . This worm hole solution is a necessary and sufficient condition for thermal transfer of heat from that prior universe to allow for graviton production under relic inflationary conditions. To model this, we use results from Lawrence B. Crowell's book on quantum fluctuations in space time which gives a model from a pseudo time component version of the Wheeler DeWitt equation, with a use of the Reinssner-Nordstrom metric to help us obtain a solution which passes through a thin shell separating two space times.

The radius of the shell, $r_0(t)$ separating the two space times is of length $l_P$ in approximate magnitude, leading to a domination of the time component for the Reissner – Nordstrom metric

$$dS^2 = -F(r) \cdot dt^2 + \frac{dr^2}{F(r)} + d\Omega^2 \qquad (1)$$

This has

$$F(r) = 1 - \frac{2M}{r} + \frac{Q^2}{r^2} - \frac{\Lambda}{3} \cdot r^2 \xrightarrow[T \to 10^{32} \, Kelvin \sim \infty]{} -\frac{\Lambda}{3} \cdot (r = l_P)^2 \qquad (2)$$

This assume that the cosmological vacuum energy parameter has a temperature dependence as outlined by Park (2003) leading to if

$$\frac{\partial F}{\partial r} \sim -2 \cdot \frac{\Lambda}{3} \cdot (r \approx l_P) \equiv \eta(T) \cdot (r \approx l_P) \qquad (3)$$

a wave functional solution to a Wheeler De Witt equation bridging two space times has a character similar to that of a form exists between these two space times with 'instantaneous' transfer of thermal heat and energy given by

$$\Psi(T) \propto -A \cdot \{\eta^2 \cdot C_1\} + A \cdot \eta \cdot \omega^2 \cdot C_2 \qquad (4)$$

This has $C_1 = C_1(\omega, t, r)$ as a pseudo cyclic and evolving function in terms of frequency, time, and spatial function, with the same thing describable about $C_2 = C_2(\omega, t, r)$ with $C_1 = C_1(\omega, t, r) \neq C_2(\omega, t, r)$. The upshot of this is that a thermal bridge between a shrinking prior universe, collapsing to a singularity, and an expanding universe expanding from a singularity exits, with an almost instantaneous transfer of heat with terms dominated by $\eta(T)$ exits, and is forming a necessary and sufficient condition for the thermal heat flux leading to the relic graviton burst delineated later on.

## Theorem 1 The following are equivalent

1. **There exists a Reisnner-Nordstrom Metric with -F(r) dt² dominated by a cosmological vacuum energy term, $(-\Lambda/3)$ times $dt^2$, for early universe conditions in the time range less than or equal to Planck's time $t_P$**
2. **A solution for a pseudo – time dependent version of the Wheeler De Witt equation exists with a wave function $\Psi(r,t,T)$ forming a wormhole bridge between two universe domains, with $\Psi(r,t,T) = \Psi(r,-t,T)$ for a region of space time before signal causality discontinuity, and for times $|t| < t_P$**

3. The heat flux dominated vacuum energy value given by $\Psi(r,t,T)$ contributes to a relic graviton burst, in a region of time less than or equal to Planck's time $t_P$

The proof of Theorem 1 will await further developments being presented in the text.

## II. Presenting evidence for causal discontinuity due to the transferal of thermally based vacuum energy as implied by the worm hole solution to the Wheeler De Witt equation.

Begin first by presenting a version of the Friedmann equation given by Peter Frampton.[13]

$$(\dot{a}/a)^2 = \frac{8\pi G}{3} \cdot [\rho_{rel} + \rho_{matter}] + \frac{\Lambda}{3} \qquad (5)$$

With, for small scale factor values

$$\int da \frac{1}{\left[a^4(t) + \frac{8\pi}{\Lambda}\left[(\rho_{rel})_0 a_0^4 + (\rho_m)_0 a_0^3 a(t)\right]\right]^{1/2}} \equiv \sqrt{\frac{\Lambda}{3}} \cdot \int dt \qquad (6)$$

We get the following polynomial expression for scale factors near the big bang itself assuming $u = a^{-1}$

$$u^9 + A_1 \cdot \frac{u^8}{a_0} + A_2 \cdot \frac{u^7}{a_0^2} - A_3 \cdot \left(\frac{\Lambda}{8\pi}\right) \cdot \frac{u^5}{a_0^4} - A_4 \cdot \left(\frac{\Lambda}{8\pi}\right) \cdot \frac{u^4}{a_0^5} + A_5 \cdot \left(\frac{\Lambda}{8\pi}\right)^2 \cdot \frac{u^1}{a_0^8} + A_6 \cdot \left(\frac{\Lambda}{8\pi}\right)^2 \cdot \frac{t}{a_0^9} \cong 0 \quad (7)$$

We could go considerably higher in polynomial roots of Eqn. (7) above, depending upon the degree of accuracy we wished to obtain. This truncation so picked above is assuming a non infinite value of $u = a^{-1}$, as well as a non zero value and non infinite value for the $\Lambda$ term. In doing so, we would obtain an extremely non standard evolution for the scale factor, assuming when we do so that

$$A_1 = \frac{9}{4} \cdot \frac{(\rho_m)_0}{(\rho_{rel})_0} \quad A_2 = \frac{(\rho_m)_0^2}{(\rho_{rel})_0^2}, A_3 = \frac{1/5}{(\rho_{rel})_0^1}, A_4 = \frac{(\rho_m)_0 \cdot (1/4)}{(\rho_{rel})_0^2}, A_5 = \frac{1}{(\rho_{rel})_0^2}, A_6 = \frac{1}{(\rho_{rel})_0^2} \cdot \sqrt{\frac{\Lambda}{3}} \quad (8)$$

We argue that the existence of such a non linear equation for early universe scale factor evolution introduces a de facto 'information' barrier between a prior universe which as we argue can only include thermal bounce input to the new nucleation phase of our present universe. To see this, we can turn to Dr. Dowker's paper on causal sets [1] which require the following ordering with a relation $\prec$, where we assume that initial relic space time is replaced by an assembly of discrete elements, so as to create, initially, a partially ordered set $C$

(1) If $x \prec y$, and $y \prec z$, then $x \prec z$

(2) If $x \prec y$, and $y \prec x$, then $x = y$ for $x, y \ \varepsilon \ C$

(3) For any pair of fixed elements $x$ and $z$ of elements in $C$, the set $\{y \mid x \prec y \prec z\}$ of elements lying in between x and z is finite

Items (1) and (2) give us that we have $C$ as a partially ordered set and the third item permits local finiteness. This when combined with as a model for how the universe evolves via a scale factor equation permits us to write, after we substitute $a(t^*) < l_P$ for $t^* < t_P =$ Planck time, and $a_0 \equiv l_P$, and $a_0/a(t^*) \equiv 10^\alpha$ for $\alpha \gg 0$ into a discrete equation model of Eqn (5) leads to

**Theorem 2** **Using the Friedmann equation for the evolution of a scale factor $a(t)$, we have a non partially ordered set evolution of the scale factor with evolving time, thereby implying a causal discontinuity**

**Proof** :

$$\left[\frac{a(t^* + \delta t)}{a(t^*)}\right] - 1 < \frac{(\delta t \cdot l_P)}{\sqrt{\Lambda/3}} \cdot \left[1 + \frac{8\pi}{\Lambda} \cdot \left[(\rho_{rel})_0 \cdot 10^{4\alpha} + (\rho_m)_0 \cdot 10^{3\alpha}\right]\right]^{1/2} \xrightarrow[\Lambda \to \infty]{} 0 \quad (9)$$

**So in the initial phases of the big bang, with a very large vacuum energy, we obtain the following relation which violates (signal) causality. This for any given fluctuation of time in the 'positive' direction**

$$\left[\frac{a(t^* + \delta t)}{a(t^*)}\right] < 1 \quad (10)$$

We argue that the existence of such a violation of a causal set arrangement in the evolution of a scale factor argues for a break in information propagation from a prior universe, to our present universe. This neatly compliments the transferal of an extremely large temperature based vacuum energy via a worm hole. This worm hole transferal from a prior universe to relic conditions in today's universe is itself an argument for causal signal breakage since it would be effectively a faster than light propagation from a prior collapsing universe, to today's universe

## III. Seth Lloyd's universe as a quantum computer model, with modifications

We make use of the formula given by Seth Lloyd in arXIV [8] which related the number of operations the 'Universe' can 'compute' during its evolution. Seth Lloyd uses the idea he attributed to Landauer to the effect that the universe is a physical system which has information being processed over its evolutionary history. Lloyd also makes reference to a prior paper where he attributes an upper bound to the permitted speed a physical system can have in performing operations in lieu of the Margolis/ Levitin theorem, with a quantum mechanically given upper limit value (assuming E is the average energy of the system above a ground state value), obtaining a **first limit** with regards to a quantum mechanical average energy bound value of

$$[\#operations/\sec] \leq 2E/\pi\hbar \quad (11)$$

The **second limi**t to this number of operations is strictly linked to entropy due to considerations as to limits to memory space. Via what Lloyd writes as

$$[\#operations] \leq S(entropy)/(k_B \cdot \ln 2) \quad (12)$$

The **third limit** as to strict considerations as to a matter dominated universe relates number of allowed computations / operations within a volume for the alleged space of a universe, making the identification of this space time volume as $c^3 \cdot t^3$, with c the speed of light, and t an alleged time/ age for the universe. We further identify $E(energy) \sim \rho \cdot c^2$, with $\rho$ as the density of matter, and $\rho \cdot c^2$ as the energy density/ unit volume. This leads to

$$[\#operations/\sec] \leq \rho \cdot c^2 \times c^3 \cdot t^3 \tag{13}$$

We then can write this, if we identify $\rho \sim 10^{-27} kil/meter^3$ and time as approximately $t \sim 10^{10}$ years as leading to a present upper bound of

$$[\#operations] \approx \rho \cdot c^5 \cdot t^4 \leq 10^{120} \tag{14}$$

Seth Lloyd further refines this to read as follows

$$\#operations = \frac{4E}{\hbar} \cdot \left(t_1 - \sqrt{t_1 t_0}\right) \approx \left(t_{Final}/t_P\right) \leq 10^{120} \tag{15}$$

We assume that $t_1$ = final time of physical evolution, whereas $t_0 = t_P \sim 10^{-43}$ seconds and that we can set an energy input via assuming in early universe conditions that $N^+ \neq \varepsilon^+ \ll 1$, and $0 < N^+ < 1$, so that we are looking at a graviton burst supplied energy value along the lines of

$$E = (V_{4-Dim}) \cdot \left[\rho_{Vac} = \frac{\Lambda}{8\pi G}\right] \sim N^+ \cdot \left[\rho_{graviton} \cdot V_{4-vol} \approx \hbar \cdot \omega_{graviton}\right] \tag{16}$$

Furthermore, if we use the assumption that the temperature is within the given range of $T \approx 10^{32} - 10^{29}$ Kelvin initially, we have that a Hubble parameter defined along the route specified by Seth Lloyd. This is in lieu of time $t = 1/H$, a horizon distance defined as $\approx c/H$, and a total energy value within the horizon as

$$\text{Energy (within the horizon)} \approx \rho_c \cdot c^3 / (H^4 \cdot \hbar) \approx 1/(t_P^2 \cdot H) \tag{17}$$

And this for a Horizon parameter Seth Lloyd defines as [8]

$$H = \sqrt{8\pi G \cdot [\rho_{crit}]/3 \cdot c^2} \tag{18}$$

And a early universe

$$\rho_{crit} \sim \rho_{graviton} \sim \hbar \cdot \omega_{graviton}/V_{4-Vol} \tag{19}$$

Then [14]

$$\begin{aligned}\#operations &\approx 1/[t_P^2 \cdot H] \approx \sqrt{V_{4-Vol}} \cdot t_P^{-2} / \sqrt{8\pi G \hbar \omega_{graviton}/3c^2} \\ &\approx [3\ln 2/4]^{4/3} \cdot [S_{Entrophy}/k_B \ln 2]^{4/3}\end{aligned} \tag{20}$$

**Theorem 3**: **The number of allowed operations in the evolution of the universe specifies a relationship between an evaluated volume for space time, and upper limits of released relic graviton frequencies**

**Proof:** See Eqn (20) above

# IIIa. Investigation of entropy within the z=1100 red shift CMB barrier

So far we have argued as to the existence of a high level of entropy within what could be a large volume of space relative to the initial volume present in vacuum nucleation of an initially very high energy density state of matter-energy. We shall endeavor to give more specifics as to the relationship between entropy density, overall entropy, changes in volume, as well as changes in time which we believe govern the spatial regime in which gravitons are initially coupled to matter. Key to this is asserting that the entropy density is scaled in early Universe conditions as an initial value of entropy density at a proper time proportional to Planck's time interval of about ten to the minus 43 seconds in magnitude times the ration of initial proper time, over proper time at a later stage in cosmological evolution. . Whereas the overall entropy is the entropy density, times a space time volume specified at a given proper time. For those who wish to understand what we are referring to, we define first a de facto distance to any comoving observer, where $D_G(t_{present})$ is the distance $D_{now}$ to galaxy G *now*, while a(t) is a universal scale factor that applies to all comoving objects. From its definition we see that $a(t_{present}) = 1$ so we get a generalized distance relationship

$$D_G(t) = a(t) * D_G(t_{present}) \qquad (21)$$

This is assuming a large red shift value, where we define the red shift value $Z$ via

$$1+z = \text{sqrt}[(1+v/c)/(1-v/c)] \qquad (22)$$

The proper time so referred to, on small scales, is based on the concept of volume, while for large scales the usual definition of length is applied. This is given in arXiv:gr-qc/0102088,[15]. Needless to say for area/volume about the big bang, proper time is proportional to the Planck time interval $\propto 10^{-43}$ seconds. This leads to us following scaling of entropy density, as given by Seibert (and Bjorken in 1983) in 1991 as, where we define $\tau_0$ as an initial (at the point of nucleation of a vacuum state) proper time value, so we write entropy density for proper times $\tau$ greater than $\tau_0$ [16]

$$s(\tau) = s(\tau_0) \cdot \left[\frac{\tau_0}{\tau}\right] \qquad (23)$$

If we make a relationship between entropy density, and entropy itself, $S(\tau) = s(\tau_0) \cdot [V_{4-\dim}] \cdot \left[\frac{\tau_0}{\tau}\right]$ we can define the onset of early stages of entropy as growing up to a time Planck time interval $\propto 10^{-43}$ seconds via Silbert's formula of [16]

$$S(t) = k \cdot \sigma \cdot t^2, \quad 0 < t < [\lambda/2c] \approx t_P \sim 10^{-43} \text{ seconds} \qquad (24)$$

This would correspond to the Loop quantum gravity insertion of thermal energy into a present universes space time continuum with an initial relic graviton producing burst initiated by a cosmological 'constant' $\Lambda_{Park} \approx \infty \xrightarrow{Axion\ mass \to 0} \Lambda_{Barvinsky} \approx 360 \cdot m_P^2$ [17], with the lower value signifying a release of relic gravitons, and this when we are setting $m_P$ as the Planck mass, i.e. the mass of a black hole of 'radius' on the order of magnitude of Planck length $l_P \sim 10^{-35}$ centimeters in width. Hereafter, we are then setting the entropy as scaling as

$$S(t) = \langle s \rangle_{net} \cdot [V_{4-\dim}(t)], \quad t_P \sim 10^{-43} < t < t\left(at \quad z \cong 1100\right) \qquad (25)$$

This is assuming that the $\langle s \rangle_{net}$ is an average value of entropy density which would be relatively constant in the aftermath of the big bang up to the red shift barrier of Z of the order of 1100. Then afterwards, we could expect a rough scaling of entropy density according to

$$S(\tau) = s(\tau_0) \cdot [V_{4-\dim}(time)] \cdot \left[\frac{\tau_0}{\tau}\right] \qquad (26)$$

We wish now that we have described how entropy behaves via scaling arguments look at the physical inputs into this problem Note, we are making one specific identification here. That the initial growth of graviton based entropy density is in tandem with the growth of free energy with increasing temperatures in the onset of a vacuum state according to

$$\text{Free energy} \sim -\pi^2 T^4/90 \approx -s(T)^2 \qquad (27)$$

We obtained this value of free energy by associating Eqn. (27) with the free energy of a massless spin zero boson, or minus the pressure of a 'spin zero boson' state, and $[\pi^2 T^4/30]$ as the energy density of a spin zero boson gas, which can be read off as part of a finite temperature one loop full potential $V_T(\phi_C)$ for high temperature values of a scalar field given by (assuming that $V(\phi_C)$ is a one loop effective potential) when we are looking at an a early universe scalar field model of inflation for which we are looking at high temperature symmetry restoration, where we look at a critical value of the scalar field at about the time of vacuum nucleation, we call $\phi_C$ at time $t \approx 10^{-43}$ seconds.

$$V_T(\phi_C) = V(\phi_C) + \frac{\lambda}{8}T^2\phi_C^2 - s(T^2) + ... \tag{28}$$

This leads to us now considering what physical inputs should be made into the parameters of our entropy/ number of computations upper limits as specified by the given equations written up in this document

# IV. Evolution of the Vacuum energy via 4$^{th}$ and 5$^{th}$ dimensional considerations, as well as the Hartle-Hawking's wave function

First of all we need to consider if there is an inherent fluctuation in early universe cosmology which is linked to a vacuum state nucleating out of 'nothing'.
.
The vacuum fluctuation leads to production of a dark energy density which we can state is initially due to contributions from an axion wall, which is dissolved during the inflationary era. What we will be doing is to reconcile how that wall was dissolved in early universe cosmology with quantum gravity models, brane world models, and Weinberg's prediction (published as of 1972) of a threshold of 10 to the 32 power Kelvin for which quantum effects become dominant in quantum gravity models. All of this leads up to conditions in which we can expect relic graviton production which could account for the presence of strong gravitational fields in the onset of Guth style inflation, would be in line with Penrose's predictions via the Jeans inequality as to low temperature, low entropy conditions for pre inflationary cosmology.

It is noteworthy that Barvinsky et al in late 2006 [17] recently predicted a range of values of four dimensional Planck's constant between upper and lower bounds. I.e. this is a way to incorporate the existence of a cosmological constant at about a Planck's time $t_P$ with the formation of scale factors which permit the existence of definable space time metrics

. A good argument can be made that prior to Planck's time $t_P$ that conventional space time metrics, even those adapting to strongly curved space do not apply. Park et al predict an upper range of cosmological constant values far in excess of Barvinsky's prediction, and we explain the difference in terms of a thermal/ vacuum energy input into graviton production. We will henceforth investigate how this would affect the emergence of an initial state for the scale factor, in the cases where the cosmological constant is first a lower bound, and then where the cosmological constant parameter is grows far larger.

A very large cosmological vacuum energy leads to a road map for solving the land scape problem. i.e. $\Lambda_{4-\text{dim,max}}(Barvinsky) = 3m_P^2/2B = 360m_P^2$ as a peak value, after graviton production would lead to a Hartle-Hawking's universe wave function [17] of the form for $\Lambda_{4-\text{dim,max}}(Barvinsky) = 3m_P^2/2B = 360m_P^2 << \Lambda_{4-\text{dim,max}}(Park) = c_1 \cdot (T_{\max} \approx 10^{23} Kelvin)$

$$\psi_{HH}\big|_{Barvinsky} \approx \exp(-S_E) = \exp(3 \cdot \pi/2 \cdot G\Lambda) \neq 0 \tag{29}$$

Conversely Parks values for a nearly infinite cosmological constant parameter, due to high temperatures would lead to, prior to graviton production, if we pick a five dimensional $\Lambda_{5-\text{dim}} \sim -c_2/(T \approx 10^{23} Kelvin)^\beta$ for embedding a high temperature version of the Park $\Lambda_{4-\text{dim,max}}(Park)$ value in a Brane world configuration

$$\psi_{HH}\big|_{5-\text{dim}-Brane-World-density} \approx \exp(-S_E) = \exp(3 \cdot \pi/2 \cdot G\Lambda) \xrightarrow[T \to \infty]{} 0 \tag{30}$$

This allows us to make in roads into a solution to the cosmological land-scape problem discussed by Guth in 2003 at the Kalvi institute in UC Santa Barbara. I.e. why have $10^{1000}$ or so independent vacuum states as predicted by String theory?[18]

If one looks at the range of allowed upper bounds of the cosmological constant, we have that the difference between what Barvinsky et al in late 2006 predicted, and Park's upper limit as of 2003, based upon thermal input strongly hints that a phase transition is occurring at or before Planck's time . This allows for a brief interlude of quintessence

Begin with assuming that the absolute value of the five dimensional cosmological 'constant' parameter is inversely related to temperature, i.e.

$$|\Lambda_{5-dim}| \propto c_1 \cdot (1/T^\alpha) \tag{31}$$

As opposed to working with the more traditional four dimensional version of the same. Those wishing to get specific values of the constants $c_1$ and $c_2$ are referred to look at the Beckwith (2006) [10], which phrased the release of gravitons in terms of [11]

$$\Lambda_{4-dim} \propto c_2 \cdot T^\beta \tag{32}$$

We should note that this is assuming that a release in gravitons occurs which leads to the removal of graviton energy stored contributions to this cosmological parameter

$$\Lambda_{4-dim} \propto c_2 \cdot T^\beta \xrightarrow{graviton-production} 360 \cdot m_P^2 << c_2 \cdot [T \approx 10^{32} K] \tag{33}$$

Needless to say, right after the gravitons are released one still observes a drop off of temperature contributions to the cosmological constant .Then we can write, for small time values $t \approx \delta^1 \cdot t_P, 0 < \delta^1 \leq 1$ and for temperatures sharply lower than $T \approx 10^{32} Kelvin$ [11]

$$\left(\frac{\Lambda}{|\Lambda_5|} - 1\right) \approx O\left(\frac{1}{n}\right) \sim \text{\bf To the order of } (1/n) \tag{34}$$

for quantum effects to be dominant in cosmology, with a value of critical energy we will use in setting a template for relic graviton production later on.

$$E_{critical} \equiv 1.22 \times 10^{28} eV \tag{35}$$

This is presupposing that we have a working cosmology which actually gets to such temperatures at the instance of quantum nucleation of a new universe.

## V. Graviton power burst/ where did the missing contributions to the cosmological 'vacuum energy 'PARAMETER go?

To do this, we need to refer to a power spectrum value which can be associated with the emission of a graviton. Fortunately, the literature contains a working expression as to power generation for a graviton being produced for a rod spinning at a frequency per second $\omega$, due to Fontana [19], which reportedly gives for a rod of length $\widehat{L}$ and of mass m a formula for graviton production power,

$$P(power) = 2 \cdot \frac{m_{graviton}^2 \cdot \widehat{L}^4 \cdot \omega_{net}^6}{45 \cdot (c^5 \cdot G)} \tag{36}$$

Note here that we need to say something about the contribution of frequency needs to be understood as a mechanical analogue to the brute mechanics of graviton production. We can view the frequency $\omega_{net}$ as an input from an energy value, with graviton production number (in terms of energy) as given approximately via an integration of Eqn. (37) below, $\hat{L} \propto l_P$ mass $m_{graviton} \propto 10^{-60} kg$. It also depends upon a huge number of relic gravitons being produced, due to the temperature variation so proposed. [10,11]

$$\langle n(\omega) \rangle = \frac{1}{\omega \left( net \ value \right)} \int_{\omega 1}^{\omega 2} \frac{\omega^2 d\omega}{\pi^2} \cdot \left[ \exp\left(\frac{2 \cdot \pi \cdot \hbar \cdot \omega}{kT}\right) - 1 \right]^{-1} \quad (37)$$

Thus, one can set a normalized 'energy input 'as $E_{eff} \equiv \langle n(\omega) \rangle \cdot \omega \equiv \omega_{eff}$; with $\hbar \omega \xrightarrow{\hbar=1} \omega \equiv |E_{critical}|$ which leads to the following table of results, with $T^*$ being a heating up value of temperature from a brane world thermal input from a prior universe quantum bounce after a nearly zero degrees Kelvin starting point of the pre inflationary universe condition specified by Carroll.. For the sake of scaling, we will refer to $T^* \sim \frac{1}{3} \times 10^{32}$ Kelvin [14], with a peak graviton burst happening at the time where quantum gravity becomes a dominant contribution to early universe vacuum energy nucleation. I.e. this phase transition occurs in a very brief instant of cosmological time with the onset of the graviton burst being modeled at times $t << t_P \sim 10^{-43}$ .seconds. The values of N1, to N5 are partly scaled graviton burst values. The tie in of a relic graviton burst so presented with brane world models has been partly explained in the author's publication and our description of a link of the sort between a brane world effective potential and eventual Guth style inflation has been partly replicated by Sago, Himenoto, and Sasaki [20] in November 2001 where they assumed a given scalar potential, assuming that $m$ is the mass of the bulk scalar field. This permits mixing the false vacuum hypothesis of Coleman in 4 dimensions with brane world theory in five dimensions.

$$V(\phi) = V_0 + \frac{1}{2} m \phi^2 \quad (38)$$

Their model is in part governed by a restriction of their 5-dimensional metric to be of the form, with $l =$ brane world curvature radius, and $\hat{H}$ H their version of the Hubble parameter

$$dS^2 = dr^2 + (\hat{H} \cdot l)^2 \cdot dS^2_{4-dim} \quad (39)$$

I.e. if we take $k_5^2$ as being a 5 dimensional gravitational constant

$$\hat{H} = \frac{k_5^2 \cdot V_0}{6} \quad (40)$$

Our difference with Eqn. (38) is that we are proposing that it is an intermediate step, and not a global picture of the inflation field potential system. However, the paper they present with its focus upon the zero mode contributions to vacuum expectations $\langle \delta\phi^2 \rangle$ on a brane has similarities as to what we did which should be investigated further. The difference between what they did, and our approach is in their value of

$$dS^2_{4-dim} \equiv -dt^2 + \frac{1}{H^2} \cdot \left[ \exp(2 \cdot \hat{H} \cdot t) \right] \cdot dx^2 \quad (41)$$

This assumes one is still working with a modified Gaussian potential all the way through, This is assuming that there exists an effective five dimensional cosmological parameter which is still less than zero, with $\Lambda_5 < 0$, and $|\Lambda_5| > k_5^2 \cdot V_0$ so that [20]

$$\Lambda_{5,eff} = \Lambda_5 + k_5^2 \cdot V_0 < 0 \quad (42)$$

It is simply a matter of having

$$|m^2| \cdot \phi^2 << V_0 \tag{43}$$

And of making the following identification

$$\phi_{5-dim} \propto \tilde{\tilde{\phi}}_{4-dim} \equiv \tilde{\tilde{\phi}} \approx [\phi - \varphi_{fluctuations}]_{4-dim} \tag{44}$$

With $\varphi_{fluctuations}$ in Eqn. (34) is an equilibrium value of a true vacuum minimum for a chaotic four dimensional quadratic scalar potential for inflationary cosmology. This in the context of the fluctuations having an upper bound of $\tilde{\tilde{\phi}}$ (Here, $\tilde{\tilde{\phi}} \geq \varphi_{fluctuations}$). And $\tilde{\tilde{\phi}}_{4-dim} \equiv \tilde{\tilde{\phi}} - \frac{m}{\sqrt{12 \cdot \pi \cdot G}} \cdot t$, where we use $\tilde{\tilde{\phi}} > \sqrt{\frac{60}{2 \cdot \pi}} M_P \approx 3.1 M_P \equiv 3.1$, with $M_P$ being a Planck's mass. This identifies an imbedding structure we will elaborate upon later on. This will in its own way lead us to make sense of a phase transition we will write as a four dimensional embedded structure within the 5 dimensional Sundrum brane world structure and the four dimensional [10,11]

$$\begin{array}{ll} \tilde{V}_1 & \rightarrow \tilde{V}_2 \\ \tilde{\phi}(increase) \leq 2 \cdot \pi & \rightarrow \tilde{\phi}(decrease) \leq 2 \cdot \pi \\ t \leq t_P & \rightarrow t \geq t_P + \delta \cdot t \end{array} \tag{45}$$

The potentials $\tilde{V}_1$, and $\tilde{V}_2$ will be described in terms of chaotic inflationary scalar potential system. Here, $m^2 \approx (1/100) \cdot M_P^2$

$$\tilde{V}_1(\phi) \propto \frac{m_a^2(T)}{2} \cdot (1 - \cos(\tilde{\phi})) + \frac{m^2}{2} \cdot (\tilde{\phi} - \phi^*)^2 \quad - \tag{46}$$

$$\tilde{V}_2(\phi) \propto \frac{1}{2} \cdot (\tilde{\phi} - \phi_C)^2 \tag{47}$$

The transition from Eqn. (46) above to Eqn. (47) is when we have a relic graviton burst as given below which also is when we have a removal of an axion wall contribution with the axion mass term $m_a(T) \xrightarrow[Temp \rightarrow Planck\ temperature]{} \varepsilon^+ \approx$ very small value almost non existent in contribution due to temperature scaling as given below.[21]

$$m_a(T) \cong 0.1 \cdot m_a(T=0) \cdot (\Lambda_{QCD}/T)^{3.7} \tag{48}$$

We assert that we need a five dimensional brane world picture to formulate what configuration the non zero axion mass makes initially to a Sundrum initial compactified 5$^{th}$ dimensional presentation of the action integral as given in Beckwith (2006)[10,11]. As the contribution to Eqn. (48) vanishes, we see the following graviton burst.

## Table 1 How to outline the existence of a relic graviton burst

| | |
|---|---|
| N1=1.794 E-6 for $Temp = T^*$ | Power = 0 |
| N2=1.133 E-4 for $Temp = 2T^*$ | Power = 0 |
| N3= 7.872 E+21 for $Temp = 3T^*$ | Power = 1.058 E+16 |
| N4= 3.612E+16 for $Temp = 4T^*$ | Power $\cong$ very small value |
| N5= 4.205E-3 for $Temp = 5T^*$ | Power= 0 |

The outcome is that there is a distinct power spike associated with Eqn. 36 and Eqn. 37, which is congruent with a relic graviton burst. Future research objectives will be to configure the conditions via brane world dynamics leading to graviton production. A good working model as to how the cosmological constant changes in this comparison of four and five dimensional cosmological constants is provided in tables given in Beckwith arXIV physics article mentioned in the reference section..

### VI formation of the scalar field, bifurcation results

Start with Padamans's formulas[12]

$$V(t) \equiv V(\phi) \sim \frac{3H^2}{8\pi G} \cdot \left(1 + \frac{\dot{H}}{3H^2}\right) \quad (49)$$

$$\phi(t) \sim \int dt \cdot \sqrt{\frac{-\dot{H}}{4\pi G}} \quad (50)$$

If $H = \dot{a}/a$, Eqn. (50) gives us zero scalar field values at the beginning of quantum nucleation of a universe, and at the point of accelerated expansion due to the final value of the cosmological constant an accelerating value of the cosmological scale factor expansion rate. We justify this statement, by going to early universe expansion models which have $a(t_{INITIAL}) \sim e^{H \cdot t(initial)}$, as juxtaposed with a later as the present time development of the scalar factor along the lines of $a(t_{later}) \sim e^{\Lambda[present-day] \cdot t(later)}$. Both regimes, if we use them in Eqn. (50) above lead to zero values for a quintessence scalar field. But it does not stop there. We will show later that in actuality that the scalar field so considered actually likely damps out far before the CMBR barrier value of expansion when Z = 1100 about 380,000 to 400,000 years after the big bang.

**Theorem 4**: We observe a zero value of the scalar field $\phi(t)$ at the onset of the big bang, and in the present to future era of cosmological evolution.

**Proof**: Eqn (50) has the time derivative of $H = \dot{a}/a$ go to zero when the scale factor $a(t_{INITIAL}) \sim e^{H \cdot t(initial)}$, and $a(t_{later}) \sim e^{\Lambda[present-day] \cdot t(later)}$

**Lemma:** The existence of two zero values of the scalar field $\phi(t)$ at the onset, and at a later time imply a bifurcation behavior for modeling quintessence scalar fields.

# VI a. Physical argument using inflationary dynamics arguing for an initially very large cosmological Vacuum energy

How can we come up with physical conditions leading to a large cosmological constant? The easiest way to do so is to look at an argument provided by Thanu Padmanabhan [22], leading to the observed cosmological constant value suggested by Park

$$\rho_{VAC} \sim \frac{\Lambda_{observed}}{8\pi G} \sim \sqrt{\rho_{UV} \cdot \rho_{IR}} \qquad (51)$$
$$\sim \sqrt{l_{Planck}^{-4} \cdot l_H^{-4}} \sim l_{Planck}^{-2} \cdot H_{observed}^2$$

$$\Delta \rho \approx \text{a dark energy density} \sim H_{observed}^2 / G \qquad (52)$$

We can replace $\Lambda_{observed}, H_{observed}^2$ by $\Lambda_{initial}, H_{initial}^2$. In addition we may look at inputs from the initial value of the Hubble parameter to get the necessary e folding needed for inflation, according to

$$E - foldings = H_{initial} \cdot \left( t_{End\ of\ \inf} - t_{beginning\ of\ \inf} \right) \equiv N \geq 100 \qquad (53)$$
$$\Rightarrow H_{initial} \geq 10^{39} - 10^{43}$$

Leading to

$$a(End-of-\inf)/a(Beginning-of-\inf) \equiv \exp(N) \qquad (54)$$

If we set $\Lambda_{initial} \sim c_1 \cdot [T \sim 10^{32} Kelvin]$ implying a very large initial cosmological constant value, we get in line with what Park suggested for times much less than the Planck interval of time at the instant of nucleation of a vacuum state

$$\Lambda_{initial} \sim [10^{156}] \cdot 8\pi G \approx \infty \qquad (55)$$

**Theorem 5**: **The existence of inflation itself in the onset of cosmological expansion implies a huge cosmological vacuum energy in relic conditions**

**Proof** : **See Eqn (55) above, and its development from first principles in this section**

This will be in tandem with our next result which specifies a region of space which is necessary for analyzing physical processes connected with graviton production, and the initial onset of a chaotic region of space time with wildly fluctuating minimal conditions for the scale factor.

$$\Lambda_{4-\text{dim}} \propto c_2 \cdot T \xrightarrow{graviton-production} 360 \cdot m_P^2 << c_2 \cdot \left[ T \approx 10^{32} K \right] \quad (56)$$

# VI b. Randall Sundrum effective potential. Tie in with 5$^{th}$ dimensions presented

The consequences of the fifth-dimension show up in a simple warped compactification involving two branes, i.e., a Planck world brane, and an IR brane. Let's call the brane where gravity is localized the Planck brane. This construction [4,10,11] permits (assuming K is a constant picked to fit brane world requirements)

$$S_5 = \int d^4 x \cdot \int_{-\pi}^{\pi} d\theta \cdot R \cdot \left\{ \frac{1}{2} \cdot (\partial_M \phi)^2 - \frac{m_5^2}{2} \cdot \phi^2 - K \cdot \phi \cdot [\delta(x_5) + \delta(x_5 - \pi \cdot R)] \right\} \quad (57)$$

Here, what is called $m_5^2$ can be linked to Kaluza Klein "excitations" via (for a number $n > 0$)

$$m_n^2 \equiv \frac{n^2}{R^2} + m_5^2 \quad (58)$$

To build the Kaluza–Klein theory, one picks an invariant metric on the circle $S^1$ that is the fiber of the $U$ (1)-bundle of electromagnetism. This leads to construction of a two component scalar term with contributions of different signs[10,11]

$$S_5 = -\int d^4 x \cdot V_{eff} \left( R_{phys}(x) \right) \rightarrow -\int d^4 x \cdot \tilde{V}_{eff} \left( R_{phys}(x) \right) \quad (59)$$

We should briefly note what an effective potential is in this situation.
We get

$$\tilde{V}_{eff} \left( R_{phys}(x) \right) = \frac{K^2}{2 \cdot m_5} \cdot \frac{1 + \exp(m_5 \cdot \pi \cdot R_{phys}(x))}{1 - \exp(m_5 \cdot \pi \cdot R_{phys}(x))} + \frac{\tilde{K}^2}{2 \cdot \tilde{m}_5} \cdot \frac{1 - \exp(\tilde{m}_5 \cdot \pi \cdot R_{phys}(x))}{1 + \exp(\tilde{m}_5 \cdot \pi \cdot R_{phys}(x))} \quad (60)$$

This above system has a metastable vacuum for a given special value of $R_{phys}(x)$. Start with

$$\Psi \propto \exp(-\int d^3 x_{space} d\tau_{Euclidian} L_E) \equiv \exp\left(-\int d^4 x \cdot L_E\right) \quad (61)$$

$$L_E \geq |Q| + \frac{1}{2} \cdot (\tilde{\phi} - \phi_0)^2 \{ \} \xrightarrow{Q \to 0} \frac{1}{2} \cdot (\tilde{\phi} - \phi_0)^2 \cdot \{ \} \tag{62}$$

Part of the integrand in Eqn. (61) is known as an action integral, $S = \int L \, dt$, where L is the Lagrangian of the system. Where as we also are assuming a change to what is known as Euclidean time, via $\tau = i \cdot t$, which has the effect of inverting the potential to emphasize the quantum bounce hypothesis of Sidney Coleman. In that hypothesis, $L$ is the Lagrangian with a vanishing kinetic energy contribution, i.e. $L \to V$, where $V$ is a potential whose graph is 'inverted' by the Euclidian time. Here, the spatial dimension $R_{phys}(x)$ is defined so that

$$\tilde{V}_{eff}(R_{phys}(x)) \approx \text{constant} + \tfrac{1}{2} \cdot (R_{phys}(x) - R_{critical})^2 \propto \tilde{V}_2(\tilde{\phi}) \propto \frac{1}{2} \cdot (\tilde{\phi} - \phi_C)^2 \tag{63}$$

And

$$\{ \} = 2 \cdot \Delta \cdot E_{gap} \tag{64}$$

We should note that the quantity $\{ \} = 2 \cdot \Delta \cdot E_{gap}$ referred to above has a shift in minimum energy values between a false vacuum minimum energy value, $E_{\text{false min}}$, and a true vacuum minimum energy $E_{\text{true min}}$, with the difference in energy reflected in Eqn. (64) above. This is akin to making the following identification of the axion contribution to a 4 dimensional potential for low temperatures $m_a^2(T \approx \varepsilon^+) \cong 100 \cdot m^2 \xrightarrow{T \to l \arg e} m_a^2(T \approx 10^{32} K) << m^2$ I.e. we look at axion walls specified by Kolb's book about conditions in the early universe (1991)[21] with his Eqn. (10.27) vanishing and collapsing to Guth's quadratic inflation. I.e. having the quadratic contribution to an inflation potential arise due to the vanishing of the axion contribution, as given below

$$V(a) = m_a^2 \cdot (f_{PQ}/N)^2 \cdot (1 - \cos[a/(f_{PQ}/N)]) \xrightarrow{T \to \infty} 0^+ \tag{65}$$

**Theorem 6: A five dimensional embedding of a scalar potential is congruent with the collapse of axion domain walls**

**Proof : See Eqn (63) plus the emergence of a dominant value for a quadratic potential as given by Eqn (65)**

# VII. Consequences of DYNAMICS of Axion Interaction with Baryonic matter, via quintessence scalar field

Bo Feng et al. [23]as of 2006 presented a phenomenological effective Lagrangian to provide an argument as to possible CPT violations in the early universe. Their model stated, specifying a non zero value to $(\partial_0 \phi)$ where $(\partial_0 \phi) \neq 0$

$$L_{eff} \sim (\partial_u \phi) \cdot A_v \cdot \begin{pmatrix} 0 & -B_x & -B_y & B_z \\ B_x & 0 & E_z & -E_y \\ B_y & -E_z & 0 & E_X \\ B_z & E_y & -E_x & 0 \end{pmatrix} \tag{66}$$

We are providing a necessary set of conditions for $(\partial_0 \phi) \neq 0$ via a first order early universe phase transition. Bo Feng and his research colleagues gave plenty of evidence via CMB style arguments as to the existence of non zero 'electric' and 'magnetic' field contributions to the left hand side of Eqn. (66) above.

In addition they state that TC and GC correlation power spectra require the violation of parity, a supposition which seems to be supported via a change of a polarization plane $\Delta \alpha$ via having incoming photons having their polarization vector of each incoming photon rotated by an angle $\Delta \alpha$ for indicating what they call 'cosmological birefringence' . The difference between their results and ours lies in the CMB limiting value of z as no larger than about 1000 to 1100, i.e. as of when the universe was just 400,000 years past the big bang. This indicates a non zero $(\partial_0 \phi) \neq 0$, but in itself does not provide dynamics as to the early evolution of a quintessence scalar field, which our manuscript out lines above. The work done by Bo Feng is still limited by the zero probability of observing relic photon production at the onset of inflation, at or about a Planck time instance, whereas what we did was to give predictions as to how quintessence in scalar fields evolves before the z = 1100 red shift

This discussion is modeled on an earlier paper on Quintessence and spontaneous Leptogenesis (baryogenesis) by [24] M. Li, X. Wang, B.Feng, and Z. Zhang which gave an effective Lagrangian, and an equation of 'motion' for quintessence which yielded four significant cases for our perusal. The last case , giving a way to reconcile the influx of thermal energy of a quantum bounce into an axion dominated initial cosmology, which lead to dissolution of the excess axion 'mass'. This final reduction of axion 'mass' via temperature variation leads to the Guth style chaotic inflationary regime.

Let us now look at a different effective Lagrangian which has some similarities to B. Feng's effective Lagrangian, Eqn. (66), but which leads to our equations of motion for Quintessence scalar fields, assuming as was in Eqn. (66) that specifying a non zero value to $(\partial_0 \phi)$ where $(\partial_0 \phi) \neq 0$ is implicitly assumed i.e.

$$L_{eff} \propto \frac{\tilde{c}}{M} \cdot (\partial_\mu \phi) \cdot J^\mu \qquad \qquad .. (67)$$

What will be significant will be the constant, $\tilde{c}$ which is the strength of interaction between a quintessence scalar field and baryonic matter. M in the denominator is a mass scale which can be either $M \equiv M_{planck}$, or $M \equiv M_{GUT}$ is not so important to our discussion, and $J^\mu$ is in reference to a baryonic 'current'. The main contribution to our analysis this paper gives us is in their quintessence 'equation of motion' which we will present, next. Note, that what we are calling $g_b$ is the degrees of freedom of baryonic states of matter, and $T$ is a back ground temperature w.r.t. early universe conditions. $H \cong 1/t(time)$ is the Hubble parameter, with time $t \propto O(t_P)$, i.e. time on the order of Planck's time, or in some cases much smaller than that.

$$\ddot{\phi} \cdot \left[1 + \frac{\tilde{c}}{M^2} \cdot \frac{T^2}{6} \cdot g_b\right] + 3 \cdot H \cdot \dot{\phi} \cdot \left[1 + \frac{\tilde{c}}{M^2} \cdot \frac{T^2}{6} \cdot g_b\right] + \left(\frac{\partial V_{axion-contri}}{\partial \phi}\right) \cong 0 \qquad (68)$$

Here I am making the following assumption about the axion contribution scalar potential system

$$V_{axion-contri} \equiv f[m_{axion}(T)] \cdot (1 - \cos(\phi)) + \frac{m^2}{2} \cdot (\phi - \phi_C)^2 \qquad (69)$$

For low temperatures, we can assume that prior to inflation, when $t << t_P$

$$f[m_{axion}(T)]_{T \approx 2^0 Kelvin} \sim ((50 - 100) \cdot m^2) \qquad (70)$$

And that right at the point where we have a thermal input with back ground temperatures at or greater than $10^{12} Kelvin$ we are observing for $0 < \varepsilon^+ << 1$ and times $t << t_P$

$$f[m_{axion}(T)]_{T \approx 10^{32} Kelvin} \sim ((\varepsilon^+) \cdot m^2) \quad (71)$$

This entails having at high enough temperatures

$$V_{axion-contri}|_{T>10^{32} Kelvin} \cong \frac{m^2}{2} \cdot (\phi - \phi_C)^2 \quad\ldots\ldots\ldots\ldots (72)$$

Let us now review the four cases so mentioned and to use them to analyze new physics

**CASE I:**

Temperature T very small, a.k.a. Carroll and Chen's suppositions (also see Penrose's version of the Jeans inequality)[25] and time less than $t_P$. This is the slow roll case, which is also true when we get to time >> $t_P$

$$\ddot{\phi} \cdot \left[1 + \frac{\tilde{c}}{M^2} \cdot \frac{T^2}{6} \cdot g_b\right] + 3 \cdot H \cdot \dot{\phi} \cdot \left[1 + \frac{\tilde{c}}{M^2} \cdot \frac{T^2}{6} \cdot g_b\right] + \left(\frac{\partial V_{axion-contri}}{\partial \phi}\right)$$

$$\xrightarrow[T \to 0^+]{} 3 \cdot H \cdot \dot{\phi} \cdot + \left(\frac{\partial V_{axion-contri}}{\partial \phi}\right) \cong 0 \quad (73)$$

## CASE II:

Temperature T very large and time in the neighborhood of $t_P$. This is NOT the slow roll case, and has $H \propto 1/t_P$. Note, which is important that the constant $c$ is not specified to be a small quantity

$$\ddot{\phi} \cdot \left[1 + \frac{\tilde{c}}{M^2} \cdot \frac{T^2}{6} \cdot g_b\right] + 3 \cdot H \cdot \dot{\phi} \cdot \left[1 + \frac{\tilde{c}}{M^2} \cdot \frac{T^2}{6} \cdot g_b\right] + \left(\frac{\partial V_{axion-contri}}{\partial \phi}\right)$$

$$\xrightarrow[T \to 10^{12} Kelvin]{} \ddot{\phi} + 3 \cdot H \cdot \dot{\phi} \cdot + \left(\frac{\tilde{c}}{M^2} \cdot \frac{T^2}{6} \cdot g_b\right)^{-1} \cdot \left(\frac{\partial V_{axion-contri}}{\partial \phi} = m^2 \cdot (\phi - \phi_C)\right) \cong 0$$

$$(74)$$

We then get a general, and a particular solution with $\phi_{general} \propto \exp(p \cdot t)$, $\phi_{particular} \equiv \phi_C$, $\phi_{Total} = \phi_{general} + \phi_{particular}$,

$$p^2 + 3 \cdot H \cdot p \cdot + \left(\frac{\tilde{c}}{M^2} \cdot \frac{T^2}{6} \cdot g_b\right)^{-1} \cdot (m^2) \cong 0$$

$$\to p \cong \left[-\frac{3H}{2} \cdot \left[2 - 4 \cdot \frac{m^2 \cdot M^2}{T^2 \cdot c \cdot g_b H}\right], -\left(6 \cdot \frac{m^2 \cdot M^2}{T^2 \cdot c \cdot g_b}\right) \approx -\varepsilon^+\right] \equiv [p_1, p_2] \quad (75)$$

$$\Rightarrow \phi_{general} \cong c_1 \cdot \exp(-|p_1| \cdot t) + c_2 \cdot \exp(-(|p_2| \approx \varepsilon^+) \cdot t)$$

$$\phi_{Total} = \phi_{general} + \phi_{particular} \cong \phi_C + \varepsilon_1 \cdot \phi_{initial\ value} + H.O.T. \text{ , where } \varepsilon_1 < 1 \quad (76)$$

## CASE III:

Temperature T very large and time in the neighborhood of $t_P$. This is not the slow roll case, and has $H \propto 1/t_P$. Note, which is important that the constant $c$ is specified to be a small quantity. We get much the same analysis as before except the higher order terms (H.O.T.) do not factor in

$$\phi_{Total} = \phi_{general} + \phi_{particular} \cong \phi_C + \varepsilon_1 \cdot \phi_{initial\ value} \text{ , where } \varepsilon_1 < 1 \quad (77)$$

**Case IV:**

Temperature T not necessarily large but on the way of becoming large valued, so the axion mass is not negligible, YET, and time in the neighborhood of $t_P$. This is NOT the slow roll case, and has $H > H_{t=t_P} \propto 1/t_P$. Begin with making the following approximation to the Axion dominated effective potential

$$V_{axion-contri} \equiv f[m_{axion}(T)] \cdot (1 - \cos(\phi)) + \frac{m^2}{2} \cdot (\phi - \phi_C)^2 \Rightarrow$$

$$\left(\frac{\partial V_{axion-contri}}{\partial \phi}\right) \xrightarrow{Temperature\ getting\ larger} \tag{78}$$

$$f[m_{axion}(T)] \cdot \frac{\phi^5}{125} - f[m_{axion}(T)] \cdot \frac{\phi^3}{6} + \left[(m^2 + f[m_{axion}(T)]) \cdot \phi - m^2 \phi\right]$$

Then we obtain

$$\ddot{\phi} + 3 \cdot H \cdot \dot{\phi} + \left(\frac{\tilde{c}}{M^2} \cdot \frac{T^2}{6} \cdot g_b\right)^{-1} \cdot \left(\frac{\partial V_{axion-contri}}{\partial \phi}\right) \cong 0 \tag{79}$$

This will lead to as the temperature rises we get that the general solution has definite character as follows

$$p^2 + 3 \cdot H \cdot p + \left(\frac{\tilde{c}}{M^2} \cdot \frac{T^2}{6} \cdot g_b\right)^{-1} \cdot (m^2) \cong 0$$

$$\rightarrow p \cong \left[-\frac{3H}{2} \cdot \left[1 \pm \sqrt{1 - \frac{6 \cdot M^2}{3 \cdot T^2 \cdot c \cdot g_b H} \cdot (m^2 + f[m_{axion}(T)])}\right]\right] \equiv [p_1, p_2]$$

$$\Rightarrow \phi_{general} \cong c_1 \cdot \exp(p_1 \cdot t) + c_2 \cdot \exp(p_2 \cdot t) \tag{80}$$

$$\propto [\phi(real) + i \cdot \phi(imaginary)]\ iff\ [m_{axion}(T)]\ large$$

$$\propto [\phi(real)]\ iff\ [m_{axion}(T)]\ small$$

This is in tandem with strikingly *non linear behavior* of the scalar field itself. As can be viewed from analysis. As the axion mass disappears, we go to how this compares with earlier attempts to analyze the cosmological " constant" parameter , which involves one loop approximations to effective potential calculations as given in the Cambridge university publication " General Relativity - An Einstein Centenary Survey "[26].

**Theorem 7: Scalar fields due to quintessence, with the rise of temperature associated with vacuum energy rising, tend to a real valued result, which damps out rapidly**

**Proof: See Eqn (80) above, and theorem 4.**

This above is a result we cam compare with the stress energy analogy given below, allowing us to make geometrical arguments similar in spirit to the approximations used for the above quintessence scalar fields.

# IX. Consequences of comparing stress energy tensor contributions to vacuum energy density states

The direct physical analogy we are referring to is one similar to QED, where we initially start off with a vacuum to vacuum matrix treatment of the stress tensor $T^{u,v}$ including the gravitational field via [27] the analogy

$$\langle out, vac | T^{u,v} | in, vac \rangle = -2i \frac{\delta}{\delta g_{u,v}} \langle out, vac | in, vac \rangle \qquad (81)$$

This is similar to what is obtain in QED, via a four current analogy seen in QED

$$\langle out, vac | j^u | in, vac \rangle = -2i \frac{\delta}{\delta A_u} \langle out, vac | in, vac \rangle \qquad (82)$$

The problem in all of this is that a treatment of any averaging of the stress-energy tensor has

$$\langle out, vac | T^{u,v} | in, vac \rangle = \langle T^{u,v} \rangle \xrightarrow[geometry \to Flat-space]{} 0 \qquad (83)$$

And that if we take the analogy of the Casmir parallel plate capacitors with, for $f(a) \approx A/a^4$, with $A$ set as a universal constant, and $a =$ separation distance between the Casmir plates

$$\langle T^{u,v} \rangle = f(a) \times diag[-1,1,1,-3] \xrightarrow[u,v \to 0]{} \langle T^{0,0} \rangle \equiv \rho_{Vac} \sim -A/a^4 \qquad (84)$$

How to reconcile this with the aforementioned brane theory values and what is known about singularities, we come up with the obvious conclusion. That there is more to the problem than what Eqn. (84) gives, which is in magnitude a de facto cosmological vacuum energy. Our work is a considerable elaboration upon what is given by Eqn. (84) above.

## IXa : Conclusion of a proof for the 1st theorem

**Proof**: **A non-linear equation for the Reisnner-Nordstrom metric, as dominated by huge vacuum energy values gives us an idea of what we can expect, as seen in Eqn (2) above. First, we have that Crowell presents the solution to a time dependent solution of the Wheeler De Witt equation. The relic graviton result is a by product of the same thermal flux as argued in the domination of the function given by Eqn. (2) . We will present the TFAE as appendix 1 of this document,**

## X. Important and novel results.

We have argued that the existence of such a non-linear equation for early universe scale factor evolution introduces a *de facto* 'information' barrier between a prior universe which as we argue can only include thermal bounce input to the new nucleation phase of our present universe, which in turn would argue in favor of maximum entropy, which in turn confirms the Seth Lloyd ten to the 120 power bit computing limit as mentioned above, as a model for how the universe evolves

.

# X b. Conclusions

1. Preliminary analysis suggests that we have a de facto pitch fork bifurcation of the scalar field (which is for) associated with dark energy production. We show via analytical methods that our model of Quintessence is for a very short term phenomenon.

2. We (likely) have a highly (pronounced) non linear dynamic regime for the initial evolution of the scale factor. This is due to thermal input from a large cosmological constant, which would as we go to inflation reduce to a constant value in line with and in agreement with what we observe in today's expanding universe.

3. Relic graviton production would be in tandem with a (causal) barrier of blockage of information flow from a prior universe to what we observe today.

4. If we have such non-linear behavior of the scale factor, leading to a causal information barrier between different cycles of an oscillating universe, we can now look at a new way to analyze the Wheeler – De Witt eqn; first for the case of a traditional scale factor treatment, and (then) alternatively in terms of gravitons plus the scale factor expression in terms of harmonic oscillators.

## Xc. (Suggestion for future) Additional research work with respect to modification of the Wheeler De Witt equation to take into account graviton production.

1. The 1$^{st}$ such equation is traditionally give, as stated by Kolb and Turner in their book on the early universe, as

$$\left[\frac{\partial^2}{\partial a^2} - 9\cdot\frac{\pi^2}{4G^2}\cdot\left[a^2 - \frac{\Lambda_{effective}}{3}\cdot a^4\right]\right]\cdot \Psi(a) = 0 \qquad (85)$$

As mentioned in the reference "The Early Universe, by Kolb and Turner this wave functional can be done described in terms of combination of Airy type solutions. These solutions though would not be particularly helpful when we examine what happens with widely varying $\Lambda_{effective}$ values placed in Eqn. (85) above.

2. We have looked at the influx of Gravitons, in early universe conditions. As of 1991, a SHO model type de composition of such an equation was derived by H.F. Dowker, and R. Laflamme [28] This is written up in terms of both a scalar evolution equation, and spin two graviton as follows:

$$\left[\left[\frac{\partial^2}{\partial \bar{a}^2} - \bar{a}^2 - \lambda\right] - \sum_n \left[\frac{\partial^2}{\partial d_n^2} - n^2 d_n^2 + \lambda_n\right]\right] \cdot \Psi(\bar{a}, \{d_n^\pm\}) = 0 \tag{86}$$

This is in the form of two SHO equations, with one wave functional, with the first term in parenthesis with respect to evolution of a scale factor, and the second in terms of graviton modes, as given in an index n, above. The solution of the above equation, as written up by Dowker, et al is a product of a Hermite polynomial plus other terms, along the lines of

$$\Psi(\bar{a}, \{d_n^\pm\}) = \widetilde{H}_p(\bar{a}) \cdot \exp(-\bar{a}^2/2) \cdot \prod_{n,r=\pm} \psi_{2p_n^r}^n (nd_n^{r2}) \tag{87}$$

The first term in the right hand side of Eqn (87) above is a Hermite polynomial of degree p, and the 2$^{nd}$ part, i.e. the product wave function expression is for the graviton wave functional. We intend to considerably elaborate upon this approach, taking a discretized version of the first equation to obtain a specific graviton linkage to scale factor evolution.

# Appendix 1 : Showing that T.F.A.E. in Theorem 1

**First, 1 implies 2, in Theorem 1**
We begin with

$$dS^2 = -F(r) \cdot dt^2 + \frac{dr^2}{F(r)} + d\Omega^2 \tag{1}$$

As well as look at $\frac{\partial F}{\partial r} \sim -2 \cdot \frac{\Lambda}{3} \cdot (r \approx l_P) \equiv \eta(T) \cdot (r \approx l_P) \Rightarrow$ domination of a wavefunctional $\Psi$ via a $-2 \cdot \frac{\Lambda}{3} \equiv \eta(T)$ due to

$$\Psi(T) \propto -A \cdot \{\eta^2 \cdot C_1\} + A \cdot \eta \cdot \omega^2 \cdot C_2 \tag{2}$$

**Next , 2 implies 3, in Theorem 1**
**High value of** $-2 \cdot \frac{\Lambda}{3} \equiv \eta(T)$ is due to a very high temperature which is transferred into the set up used to define inputs into Eqn (36) and Eqn (37) which results in the power burst shown in the 1$^{st}$ table.
**Next , 3 implies 1, in Theorem 1.**
**The graviton burst so identified is a way to get**

$$V(a) = m_a^2 \cdot (f_{PQ}/N)^2 \cdot (1 - \cos[a/(f_{PQ}/N)]) \xrightarrow[T \to \infty]{} 0^+ \tag{3}$$

**This forms a necessary condition for forming a removal of the embedding condition of 5$^{th}$ dimensional structure used to get the Hartle-Hawking wave functional to move from** $|\Lambda_{5-\dim}| \propto c_1 \cdot (1/T^\alpha)$ **to**

$$\Lambda_{4-\dim} \propto c_2 \cdot T^\beta \xrightarrow[graviton-production]{} 360 \cdot m_P^2 << c_2 \cdot [T \approx 10^{32} K] \tag{4}$$

**This implies a Hartle-Hawking value**

$$\psi_{HH}\big|_{5-\dim-Brane-World-density} \approx \exp(-S_E) = \exp(3 \cdot \pi/2 \cdot G\Lambda) \xrightarrow[T \to \infty]{} 0 \tag{5}$$

**Leading to**

$$\psi_{HH}\big|_{Barvinsky} \approx \exp(-S_E) = \exp(3 \cdot \pi/2 \cdot G\Lambda) \neq 0 \tag{6}$$

**This pre supposes a very high temperature flux with a maximum temperature of the order of** $T \approx 10^{32}\,^0 K$ **which is where we begin part 1 of Theorem 1.**

# Bibliography

[1] Dowker,H. F.,"Causal sets and the deep structure of spacetime", arXIV gr-qc/0508109v1 26 Aug 2005

[2] Ashtekar, A., Pawlowski, T. and Singh, P., "Quantum nature of the Big Bang," *Phys.Rev.Lett.* **96**, 141301 (2006).

[3] Crowell, L, " Quantum Fluctuations of Space Time ", World Scientific Series in Contemporary Chemical Physics, Vol 25, World Scientific, PTE, LTD, 2005, Singapore

[4] Sundrum, R., "Extra Dimensions," SLAC Summer Institute: Gravity in the Quantum World and the Cosmos, http://www-conf.slac.stanford.edu/ssi/2005/lec_notes/Sundrum1/sundrum1.pdf (2005)

[5] Buusso, R., Randall, L. , "Holographic Domains of Anti-de Sitter Space", arXIV hep-th/0112080 10 Dec 2001

[6] Carroll, S. and Chen, J. , "Does Inflation Provide Natural Initial Conditions for the Universe", arXIV gr-qc/0505037 v1 9 May 2005

[7] Weinberg , S. "Gravitation and Cosmology: Principles and Applications of the General theory of Relativity", , John Wiley & Sons, Inc. New York, 1972

[8] Lloyd , S. , "Computational capacity of the universe" , arXIV quant-ph/0110141 vol 1 24 Oct 2001

[9] Park, D.K., Kim, H., and Tamarayan, S., "Nonvanishing Cosmological Constant of Flat Universe in Brane world Senarios," http://arxiv.org/abs/hep-th/0111081 (2002).

[10] Beckwith, A.W., "How a Randall-Sundrum brane-world effective potential influences inflation physics", **AIP Conf.Proc.880:1170-1180,2007, and** arXIV physics/0610247

[11] Beckwith, A.W.,**"**Does A Randall-Sundrum Brane World Effective Potential Influence Axion Walls Helping to Form a Cosmological Constant Affecting Inflation?" http://www.citebase.org/cgi-bin/citations?id=oai:arXiv.org:gr-qc/0603021 (2006**)**

[12] Padmanabhan, T., "An Invitation to Astro physics", World Scientific Publishing Co. Pte. Ltd, Singapore, 2006

[13] Frampton, P, Baum, L, "Turnaround in Cyclic Cosmology ",Phys.Rev.Lett. 98 (2007) 071301 , arXiv:hep-th/0610213v2

[14] Beckwith, A.W., "Using quantum computing models for graviton communication/information processing in cosmological evolution", arXIV physics/ 0706.0842

[15] Filk, T., "Proper time and Minkowski structure on causal graphs", Class.Quant.Grav. 18 (2001) 2785-2796, http://arxiv.org/abs/gr-qc/0102088

[16] Seibert, S. "Initial Entropy Generation in Ultra relativistic Nuclear Collisions" Phys Rev Lett, Vol 67, Number 1, pp 12-13

[17] Barvinsky , A.O., Kamenschick, A. Yu, "Thermodynamics from Nothing: Limiting the Cosmological Constant Landscape", A.0 , Phy Rev D **74** 121502 (Rapid communications)

[18] Guth. A.H., "Eternal Inflation," http://online.itp.ucsb.edu/online/strings_c03/guth/pdf/KITPGuth_2.pdf (2003).

[19] Fontana, G., "Gravitational Wave Propulsion " CP746, Space Technology and Applications International Forum- STAIF 2005, Edited by M. S. El-Genk @ 2005 American Institute of Physics

[20] Sago, Y. Himemoto, and Sasaki, M. ,"Quantum fluctuations in brane-world inflation without an inflation on the brane", arXIV gr-qc/ 0104033 v3 9 Nov 2001

[21] Kolb, E.W. , and Turner, M.S. ," The Early universe", West view Press, 1990

[22] Padmanabhan, T. on pp 175-201 of "100 years of Relativity, Space-Time structure: Einstein and beyond", with Abhay Ashentekar, as editor, World Press scientific, PTE. Ltd. (2005), New Jersey, USA

[23] Feng, B, Li, M.,Xia, M. , Chen, Z , and Zhang, X. " Searching for CPT Violation with Cosmic Microwave Background Data from WMAP and Boomerang", arXIV 0601095 v3 8 June, 2006

[24] Li, M. ,Wang, X., Feng, B., and Zhang, X., "Quintessence and spontaneous Leptogenesis", arXIV hep-ph/0112069 Decmeber 5, 2001

[25] Penrose, R. 'The emperors New Mind', Ch 7, (Oxford: Oxford university press, 1989), and references therein

[26] See different references within "General relativity, An Einstein Centenary Survey"edited by Hawking, S.W, and Israel, W., Cambridge University Press, 1979,

[27] DeWitt, B, pp 680-745, in "General relativity, An Einstein Centenary Survey"edited by Hawking, S.W, and Israel, W., Cambridge University Press, 1979

[28] Dowker, H.F., and Laflamme, R. ," Worm holes and Gravitons" , Nuclear Physics B366 (1991) 209-232